\documentclass[amstex,twocolumn,showpacs,floats,floatfix,superscriptaddress,aps,pra]{revtex4}
\usepackage{amssymb}
\usepackage{amsmath}
\usepackage{calc}
\usepackage{graphicx}
\usepackage{bm}

\def\be{ \begin{equation} }
\def\ee{ \end{equation} }
\def\bea{ \begin{eqnarray} }
\def\eea{ \end{eqnarray} }
\def\bse{ \begin{subequations} }
\def\ese{ \end{subequations} }

\input{tcilatex}

\begin{document}

\author{G. S. Vasilev}

\affiliation{Department of Physics, Sofia University, James Bourchier 5 blvd, 1164 Sofia,
Bulgaria}
\author{N. V. Vitanov}
\affiliation{Department of Physics, Sofia University, James Bourchier 5 blvd, 1164 Sofia,
Bulgaria}
\title{Coherent excitation of two-state system by a Lorentzian filed}
\date{\today }

\begin{abstract}
This work presents an analytic description of the coherent excitation of a
two-state quantum system by an external field with a Lorentzian temporal
shape and a constant frequency. An exact analytical solution for the
differential equation of the model is derived in terms of Confluent Heun
functions. Also a very accurate analytic approximation to the transition
probability is derived by using the Dykhne-Davis-Pechukas approach. This
approximation provides analytic expressions for the frequency and amplitude
of the probability oscillations.
\end{abstract}

\maketitle


\section{Introduction}


Models involving two-state quantum system has been extensively studied since
the early days of quantum mechanics. Nowadays, such models can be found in a
variety of problems across quantum physics, ranging from nuclear magnetic
resonance, coherent atomic excitation and quantum information to chemical
physics, solid-state physics and neutrino oscillations. Moreover, many
problems involving multiple states and complicated linkage patterns can very
often be understood only by reduction to one or more two-state problems \cite%
{Shore90}. It is known, that the two-state problem with arbitrary
time-dependent fields is related to the Riccati equation and hence a general
solution cannot be derived.

There are several exactly soluble two-state models, including the Rabi \cite%
{Rabi}, Landau-Zener \cite{LZ}, Rosen-Zener \cite{RZ}, Allen-Eberly-Hioe
\cite{AEH}, Bambini-Berman \cite{BB}, Demkov-Kunike \cite{DK}, Demkov \cite%
{Demkov} and Nikitin \cite{Nikitin} models. Because of the importance of the
two-state models, the search for analytical solutions continues \cite%
{Rostovtsev}. Due to its complexity most of these models use various special
functions to solve the particular two-state problem. If this is not
possible, there exist also methods for approximate solutions, such as
adiabatic approximations, Magnus approximation, Dykhne-Davis-Pechukas
approximation.

In the present work, we derive analytically the transition probability for a
two-state system driven by a pulsed external field with Lorentzian temporal
envelope and constant carrier frequency. The solution is expressed in terms
of confluent Heun functions. This field, for which no exact analytic
solution was known yet, is among the most important pulsed fields. Using
Delos-Thorson approach presented solution can be extended to solution of
class of models \cite{Delos-Thorson}. Despite the active research the Heun
equations and Heun function have not been well studied. Because of this the
Lorentzian model will also be investigated with the Dykhne-Davis-Pechukas
(DDP) method \cite{Dykhne,Davis76}, which involves integration in the
complex time plane, to derive a very accurate approximation to the
transition probability and the width of the excitation line profile.

This paper is organized as follows. In Sec. \ref{Sec-background} we provide
the basic equations and definitions and define the problem. In Sec. we
derive a solution for the differential equation of the model in terms of
Confluent Heun functions. In Sec. wederive the transition probability by
using the DDP method. We summarize the conclusions in Sec. \ref%
{Sec-conclusion}.


\section{Basic equations and definitions\label{Sec-background}}


\subsection{Definition of the problem}


The probability amplitudes in a two-state system $\mathbf{c}(t)=\left[
c_{1}(t),c_{2}(t)\right] ^{T}$ satisfy the Schr\"{o}dinger equation,
\begin{equation}
\text{i}\hbar \frac{\text{d}}{\text{d}t}\mathbf{c}(t)=\mathbf{H}(t)\mathbf{c}%
(t),  \label{Schrodinger-2SS}
\end{equation}%
where the Hamiltonian in the rotating-wave approximation (RWA) reads \cite%
{B.Shore} $\ $%
\begin{equation}
\mathbf{H}(t)=\tfrac{1}{2}\hbar \left[
\begin{array}{cc}
-\Delta (t) & \Omega (t) \\
\Omega (t) & \Delta (t)%
\end{array}%
\right] .  \label{H2}
\end{equation}%
The detuning $\Delta $ measures the frequency offset of the field carrier
frequency $\omega $ from the Bohr transition frequency $\omega _{0}$, $%
\Delta =\omega _{0}-\omega $. The Rabi frequency $\Omega (t)$ quantifies the
field-induced coupling between the two states. For example, for laser-atom
excitation, $\Omega (t)=-\mathbf{d}.\mathbf{E}(t)/\hbar $, where $\mathbf{d}$
is the atomic transition dipole moment and $\mathbf{E}(t)$ is the laser
electric field amplitude.

We are interested in the case when the coupling has Lorentzian shape and the
detuning is constant,
\begin{subequations}
\label{model}
\begin{eqnarray}
\Omega (t) &=&\frac{\Omega _{0}}{1+t^{2}/T^{2}},  \label{Rabi} \\
\Delta (t) &=&\mathrm{const}.  \label{detuning}
\end{eqnarray}%
Because the transition probability is an even function of $\Omega _{0}$, $%
\Delta $ and $T$, for simplicity and without loss of generality all these
constants will be assumed positive.

If the system is initially in state $|\psi _{1}\rangle$ [$c_{1}(-\infty )=1$%
, $c_{2}(-\infty )=0$], the transition probability after the interaction is
given by $\mathcal{P}=\left| c_{2}(+\infty )\right| ^{2}$; its determination
is our main concern.

We shall derive below analytic solution for this model al well as several
approximations to $\mathcal{P}$ and calculate from them the period and the
amplitude of the probability oscillations, the line shape $\mathcal{P}%
(\Delta )$ and the linewidth $\Delta _{1/2}$.


\section{Reduction to Heun equation}


Heun's differential equations, which is a second order ODE of Fuchsian type
with four regular singular points has received renewed attention, together
with its various confluent forms. It turns out the solution of the
Lorentzian model is closely related to the confluent Heun equation (CHE).
Generally speaking CHE can be obtained from the general Heun equation by
coalescing of two of the singular points by redefining certain parameters
and taking the appropriate limits. In this way two regular singular points
coalesce into one irregular point. For the Lorentzian model this irregular
singular point is located at infinity, where the initial conditions are
imposed [$c_{1}(-\infty )=1$, $c_{2}(-\infty )=0]$. Although the CHE is
relatively well studied [23]-[29], there are essential gaps in the theory
and this results to some difficulties when the initial conditions for the
Lorentzian model are imposed. In general there are several canonical forms
of the CHE. The second order differential equation for the Lorentzian model
we intend to solve can be cast to one of the equivalent forms of the CHE-so
called generalized spheroidal wave equation (GSWE). Hereafter, we will use
the standart form of the GSWE

\end{subequations}
\begin{eqnarray}
z(z-z_{0})\frac{d^{2}u}{dz^{2}}+\left( B_{1}+B_{2}z\right) \frac{du}{dz}+
\label{GSWE} \\
\left[ B_{3}-2\eta \omega (z-z_{0})+\omega ^{2}z(z-z_{0})\right] u &=&0,
\notag
\end{eqnarray}

where $B_{i\text{ }},\eta $ and $\omega $ are constants with the condition $%
\omega \neq 0.$ In the GSWE $z=0$ and $z=$ $z_{0}$ are regular singular
points with indicial exponents respectively $(0,1+B_{1}/z_{0})$ and $%
(0,1-B_{2}-B_{1}/z_{0})$. At infinity i.e. $z=\infty $, GSWE has irregular
singular point. At this point the behavior of the two independent solutions
can be derived from the Thome normal solutions \cite{Olver},\cite{Leaver}
and read

\begin{equation}
\underset{z\rightarrow \infty }{\lim }u(z)\sim e^{\pm i\omega z}z^{\mp i\eta
-B_{2}/2}\left( 1+O\left( 1/z\right) \right)  \label{GSWE-inf-sol}
\end{equation}


\subsection{Solution in term fo Heun functions}


The exact analytical solution for the Lorentzina model can be derived in
terms of confluent Heun functions. The differential equation corresponding
to Lorentzian model is given by%
\begin{equation}
\frac{d^{2}c(t)}{dt^{2}}+\left( \frac{2t}{1+t^{2}}+i\Delta \right) \frac{dc}{%
dt}+\frac{\Omega _{0}^{2}}{4(1+t^{2})^{2}}c(t)=0.  \label{DELrtz}
\end{equation}

Using the substitution%
\begin{equation*}
c(t)=e^{it\Delta }\left( \frac{t+i}{t-i}\right) ^{\Omega _{0}/4}u(t)
\end{equation*}

the solution to the Loretnzian differential equation Eq.(\ref{DELrtz}) can
be expressed in terms of local confluent Heun functions

\begin{equation*}
c(t)=e^{-it\Delta }\left( \frac{t-i}{t+i}\right) ^{\Omega _{0}/4}\left[
\left( t+i\right) ^{\Omega _{0}/2}C_{1}CHl_{1}+C_{2}CHl_{2}\right] ,
\end{equation*}

where $CHl_{1}$ and $CHl_{2}$ are the local confluent Heun functions as
follows%
\begin{eqnarray*}
CHl_{1} &=&CHl\left( 2\Delta ,\frac{\Omega _{0}}{2},\frac{\Omega _{0}}{2}%
,2\Delta ,\frac{\Omega _{0}-\Delta }{8};\frac{1-it}{2}\right)  \\
CHl_{2} &=&CHl\left( 2\Delta ,-\frac{\Omega _{0}}{2},\frac{\Omega _{0}}{2}%
,2\Delta ,\frac{\Omega _{0}-\Delta }{8};\frac{1-it}{2}\right)
\end{eqnarray*}

\section{DDP approach}


\subsection{Adiabatic basis}


For the derivation of the transition probability we shall need the adiabatic
basis, i.e. the basis of the eigenstates of the Hamiltonian (\ref{Hc}). We
summarize below the basic definitions and properties of this basis.

In terms of the mixing angle $\vartheta (t)$, defined as
\begin{equation}
\tan 2\vartheta (t)=\frac{\Omega(t)}{\Delta },\qquad
(0\leqq\vartheta(t)\leqq\frac\pi4),  \label{theta}
\end{equation}
the eigenstates of $\mathsf{H}(t)$ read
\begin{subequations}
\label{adiabatic states}
\begin{eqnarray}
|\varphi _{-}(t)\rangle &=&\cos \vartheta (t)|\psi _{1}\rangle -\sin
\vartheta (t)|\psi _{2}\rangle ,  \label{phi-} \\
|\varphi _{+}(t)\rangle &=&\sin \vartheta (t)|\psi _{1}\rangle +\cos
\vartheta (t)|\psi _{2}\rangle .  \label{phi+}
\end{eqnarray}
The time dependences of the adiabatic states $|\varphi_{-}(t)\rangle $ and $%
|\varphi _{+}(t)\rangle $ derive from the mixing angle $\vartheta (t)$,
whereas the bare (diabatic) states $|\psi _{1}\rangle $ and $|\psi
_{2}\rangle $ are stationary.

Because the Rabi frequency $\Omega (t)$ vanishes at large times, and because
$\Delta>0$, we have $\vartheta (\pm \infty )=0$; hence
\end{subequations}
\begin{subequations}
\label{diabatic-adiabatic}
\begin{eqnarray}
|\varphi _{-}(\pm \infty )\rangle &=&|\psi _{1}\rangle , \\
|\varphi _{+}(\pm \infty )\rangle &=&|\psi _{2}\rangle .
\end{eqnarray}
It follows from these relations that a transition between the diabatic
states implies a transition between the adiabatic states and vice versa.
Hence the transition probability in the adiabatic basis is equal to the
transition probability in the diabatic basis.

The energies of the adiabatic states are the eigenvalues of $\mathsf{H}(t)$,
\end{subequations}
\begin{equation}
\hbar \mathcal{E}_{\pm }(t)=\frac{\hbar }{2}\left[ \Delta \pm \sqrt{%
\Omega^{2}(t)+\Delta ^{2}}\right] .
\end{equation}
The splitting between them is given by
\begin{equation}
\hbar \mathcal{E}(t)=\hbar \mathcal{E}_{+}(t)-\hbar \mathcal{E}_{-}(t)=\hbar
\sqrt{\Omega ^{2}(t)+\Delta ^{2}}.  \label{splitting}
\end{equation}
It tends $\hbar \Delta $ as $t\rightarrow \pm \infty $ and its maximum value
$\hbar \sqrt{\Omega_0^2+\Delta^2}$ is reached when $\Omega (t)$ is maximal,
at $t=0$.

The probability amplitudes in the diabatic and adiabatic bases are connected
via the rotation matrix
\begin{equation}
\mathsf{R}(\vartheta )=\left[
\begin{array}{cc}
\cos \vartheta & \sin \vartheta \\
-\sin \vartheta & \cos \vartheta%
\end{array}
\right],
\end{equation}
as
\begin{equation}
\mathbf{c}(t)=\mathsf{R}(\vartheta (t))\mathbf{a}(t),  \label{c=Ra}
\end{equation}
where the column-vector $\mathbf{a}(t)=[a_{-}(t),a_{+}(t)]^{T}$ comprises
the probability amplitudes of the adiabatic states $|\varphi _{-}(t)\rangle $
and $|\varphi _{+}(t)\rangle $. These amplitudes satisfy the transformed Schr%
\"{o}dinger equation,
\begin{equation}
i\hbar \frac{d}{dt}\mathbf{a}(t)=\mathsf{H}_{a}(t)\mathbf{a}(t),
\label{SEq-A}
\end{equation}
where the transformed Hamiltonian is given by
\begin{eqnarray}
\mathsf{H}_{a}(t) &=&\mathsf{R}^{-1}(\vartheta (t))\mathsf{H}(t)\mathsf{R}%
(\vartheta (t))-i\hbar \mathsf{R}^{-1}(\vartheta (t))\mathsf{\dot{R}}%
(\vartheta (t))  \notag \\
&=&\hbar \left[
\begin{array}{cc}
\mathcal{E}_{-}(t) & -i\dot{\vartheta}(t) \\
i\dot{\vartheta}(t) & \mathcal{E}_{+}(t)%
\end{array}
\right] ,  \label{Ha}
\end{eqnarray}
where the overdots denote time derivatives.


\subsection{The Dykhne-Davis-Pechukas approximation\label{Sec-DDP}}



\subsubsection{Single transition point}


We shall estimate the transition probability $\mathcal{P}$ for the Gaussian
model (\ref{model}) by using the Dykhne-Davis-Pechukas (DDP) approximation.
The DDP formula \cite{Dykhne,Davis76} provides the asymptotically exact
transition probability between the adiabatic states in the adiabatic limit.
We shall use this formula to calculate the transition probability $\mathcal{P%
}$ in the original, diabatic basis because, as we discussed above, the
transition probabilities in the adiabatic and diabatic bases are equal. The
DDP formula reads
\begin{equation}
\mathcal{P} \sim e^{-2\mathrm{Im}\mathcal{D}(t_{0})},  \label{DP-1}
\end{equation}
where
\begin{equation}  \label{D}
\mathcal{D}(t_{0})=\int_{0}^{t_{0}}\mathcal{E}(t)dt.
\end{equation}
The point $t_{0}$ is called the transition point and it is defined as the
(complex) zero of the quasienergy splitting,
\begin{equation}
\mathcal{E}(t_{0})=0,  \label{Tc-def}
\end{equation}
which lies in the upper half of the complex $t$-plane (i.e., with $\mathrm{%
Im\,}t_{0}>0$).
Equation (\ref{DP-1}) gives the correct asymptotic probability for
nonadiabatic transitions provided: (i) the quasienergy splitting $\mathcal{E}%
(t)$ does not vanish for real $t $, including at $\pm \infty $; (ii) $%
\mathcal{E}(t)$ is analytic and single-valued at least throughout a region
of the complex $t$-plane that includes the region from the real axis to the
transition point $t_{0}$; (iii) the transition point $t_{0}$ is well
separated from the other quasienergy zero points (if any) and from possible
singularities; (iv) there exists a level (or Stokes) line defined by
\begin{equation}  \label{Stokes line}
{\text{Im}} \mathcal{D}(t)={\text{Im}} \mathcal{D}(t_0),
\end{equation}
which goes from $-\infty$ to $+\infty$ and passes through $t_0$.

As has been pointed out already by Davis and Pechukas \cite{Davis76}, for
the Landau-Zener model \cite{LZ}, which possesses a single transition point,
the DDP formula (\ref{DP-1}) gives the exact transition probability, not
only in the adiabatic limit but also in the general case. This amazing
feature indicates not only the relevance of the DDP approximation, but
raises an intriguing, yet unanswered question: how can a (first-order)
approximate method provide the exact solution?


\subsubsection{Multiple transition points}


In the case of more than one zero points in the upper $t$-plane, Davis and
Pechukas \cite{Davis76} have suggested, following George and Lin \cite%
{George74}, that Eq.~(\ref{DP-1}) can be generalized to include the
contributions from all these $N$ zero points $t_{k}$ in a coherent sum. This
suggestion has been later verified by Joye \emph{et al.} \cite{Joye91} and
Suominen \emph{et al.} \cite%
{Suominen91,Suominen92pra,Suominen92oc,SuominenThesis}. The generalized DDP
formula has the form
\begin{equation}  \label{DP-N}
\mathcal{P}\sim \left| \sum_{k=1}^{N}\Gamma(t_k) e^{i\mathcal{D}%
(t_{k})}\right| ^{2},
\end{equation}
where $\Gamma(t_k)$ are phase factors defined by
\begin{equation}  \label{Gamma-k}
\Gamma(t_k) = 4i\lim\limits_{t\rightarrow t_{k}}(t-t_{k})\dot{\vartheta}(t).
\end{equation}

In principle, Eq.~(\ref{DP-N}) should be used when there are more than one
transition points lying on the lowest Stokes line (the closest one to the
real axis) and should include in principle only the contributions from these
points; moreover, Eq.~(\ref{DP-N}) has been rigorously proved only for these
transition points \cite{Joye91}.

Another open question for the DDP method is the parameter range where it
applies. Strictly, the DDP approximation, being a perturbative result in the
adiabatic basis, should be valid only near the adiabatic limit. For a
Gaussian field this implies the range defined by the adiabatic condition (%
\ref{adiabatic condition Gaussian}). However, we shall see that the DDP
approximation describes very accurately the transition probability well
outside this range, virtually for any parameter values, which follows
similar earlier successes of this approximation for other models (for some
of which, as we discussed, it provides even the exact result). This accuracy
of the DDP approximation well beyond the adiabatic regime, essentially in
the entire parameter plane, is another open question.


\subsection{Transition points}


For the Lorentzian model (\ref{model}), the transition points in the upper
half-plane in terms of the dimensionless time $\tau =t/T$ are given by
\begin{equation}
\tau ^{\pm }=\pm \sqrt{\frac{-1+\sqrt{1+\alpha ^{2}}}{2}}+i\sqrt{\frac{1+%
\sqrt{1+\alpha ^{2}}}{2}},  \label{trans. points}
\end{equation}%
where
\begin{equation}
\alpha =\frac{\Omega _{0}}{\Delta }.  \label{alpha}
\end{equation}

\begin{figure}[tb]
\includegraphics[width=95mm]{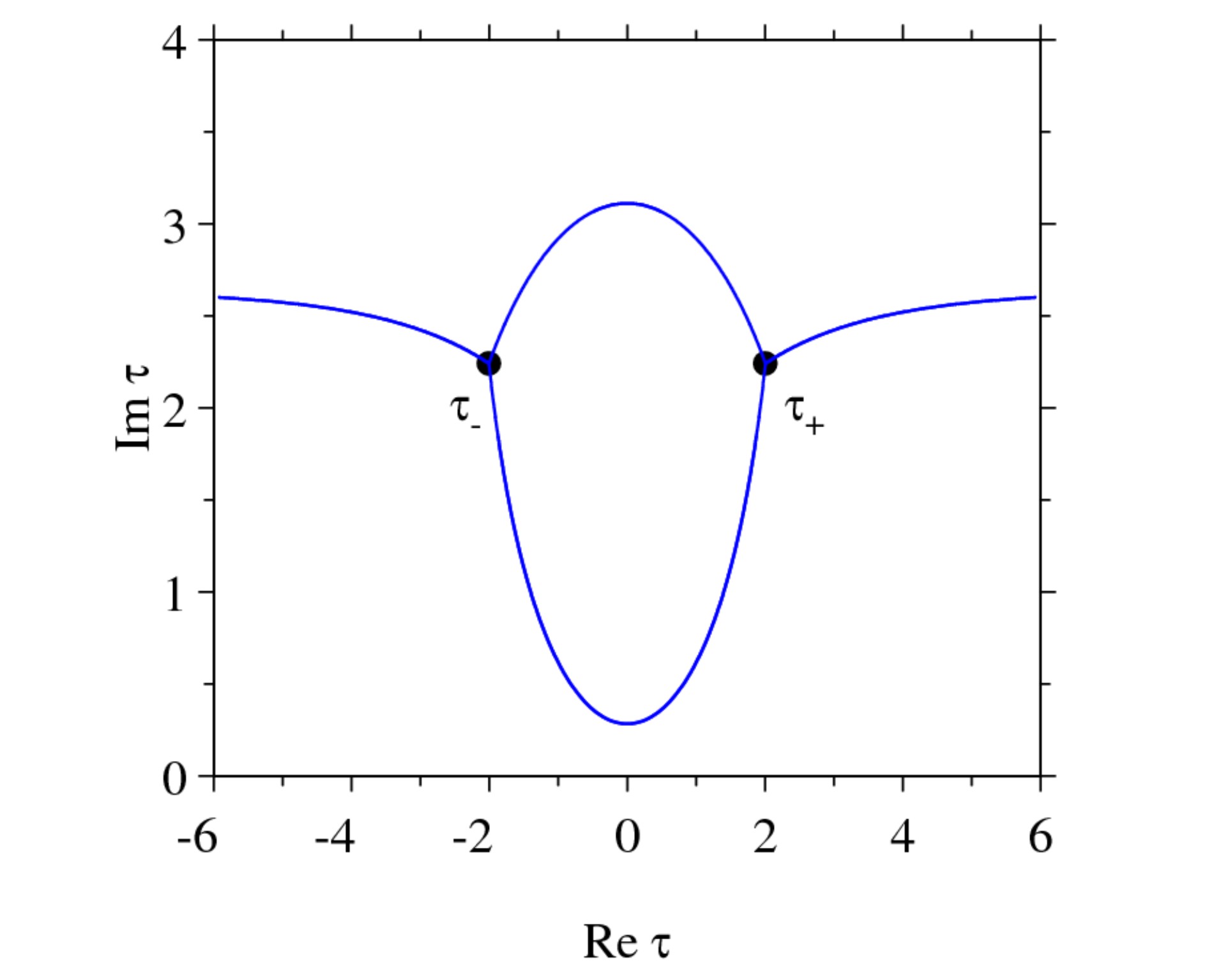}
\caption{Stokes lines for the Lorentzian model} \label{stokes}
\end{figure}

Figure \ref{stokes} displays the Stokes lines, defined by Eq.~(\ref%
{Stokes line}). For the Lorentzian model the zeroes of the eigenenergy
splitting $\mathcal{E}(t)$ are simple, and because of the presence of the
square root in $\mathcal{E}(t)$, there are three Stokes lines emerging from
each transition point \cite{Davis76}. The lowest Stokes line, which connects
$\tau ^{-}$ and $\tau ^{+}$ and extends from $-\infty $ to $+\infty $, is
the most significant one because it is used in the derivation of the DDP
approximation \cite{Davis76} and its existence validates the approximation
\cite{Joye91}.


\subsection{DDP integrals}


For the Lorentzian model we have $\left( \tau ^{-}\right) ^{\ast }=-\tau
^{+} $ and because $\mathcal{E}(\tau )$ is an even function of time, it is
easy to show that
\begin{equation}
\mathcal{D}(\tau ^{-})=-\mathcal{D}^{\ast }(\tau ^{+}),  \label{D-D}
\end{equation}%
that is ${\text{Re}}\mathcal{D}(\tau ^{-})=-{\text{Re}}\mathcal{D}(\tau
^{+}) $ and ${\text{Im}}\mathcal{D}(\tau ^{-})={\text{Im}}\mathcal{D}(\tau
^{+})$. Hence it is sufficient to calculate only one of these integrals and
we choose $\mathcal{D}(\tau ^{+})$ for this purpose.

Because the imaginary part of the DDP integral $\mathcal{D}(\tau )$ is the
same for the two transition points $\tau ^{+}$ and $\tau ^{-}$ [cf.~Eq.~(\ref%
{D-D})], these points lie on the same Stokes line, defined by Eq.~(\ref%
{Stokes line}). This Stokes line extends from $-\infty $ to $+\infty $,
which is a necessary condition for the validity of the DDP approximation
\cite{Davis76,Joye91}.

With the arguments presented above, the problem is reduced to the
calculation of the DDP integral
\begin{equation}
\mathcal{D}(\tau ^{+})=\Delta T\int_{0}^{\tau ^{+}}\frac{\sqrt{\alpha
^{2}+\left( 1+\tau ^{2}\right) ^{2}}}{1+\tau ^{2}}\,d\tau .  \label{D(Tc)}
\end{equation}

It is well known that integrals of the form $\int R[t,\sqrt{P(t)}]dt,$ where
$P(t)$ is a polynomial of third or fourth degree and $\ R$ is a rational
function can be solved in terms of three Legendre's canonical elliptic
integrals. These fundamental integrals are denoted as $E(\varphi ,k),$ $%
F(\varphi ,k)$ and $\Pi (n;\varphi ,k)$
\begin{subequations}
\begin{eqnarray}
E(\varphi ,k) &=&\int_{0}^{\varphi }\sqrt{1-k\sin ^{2}t}dt  \label{EllipticE}
\\
F(\varphi ,k) &=&\int_{0}^{\varphi }1/\sqrt{1-k\sin ^{2}t}dt
\label{EllipticF} \\
\Pi (n;\varphi ,k) &=&\int_{0}^{\varphi }1/\sqrt{(1-n\sin ^{2}t)(1-k\sin
^{2}t)}dt  \label{EllipticPi}
\end{eqnarray}%
\cite{Elliptic integrals}$.$ It is straightforward calculation to obtain the
following expression for the integral given by (\ref{D(Tc)})
\end{subequations}
\begin{equation}
\mathcal{D}(\tau ^{+})=\frac{\Delta T}{\sqrt{1-i\alpha }}\left[ \left(
-i-\alpha \right) E(\varphi ,k)+\alpha F(\varphi ,k)-i\alpha ^{2}\Pi
(n;\varphi ,k)\right] ,  \label{D(Tc)-exact}
\end{equation}

where $n,\varphi $ and $k$ are given by
\begin{equation}
\varphi =i\arcsin \text{h}\left( \frac{\tau ^{+}}{\sqrt{2+2i\alpha }}\right)
;\ k=\frac{1+i\alpha }{1-i\alpha };\;n=1+i\alpha .  \label{n-phi-k}
\end{equation}


\subsection{Transition probability}


In order to sum the contributions from the two DDP integrals we need the
factors $\Gamma _{k}$, Eq. (\ref{Gamma-k}). One finds after simple algebra
that
\begin{equation}
\Gamma (\tau ^{\pm })=\pm 1.  \label{Gamma-k-1}
\end{equation}

Now we have all the ingredients to calculate the transition probability $%
\mathcal{P}$. Collecting the results from Eqs. (\ref{D(Tc)-exact}) and (\ref%
{Gamma-k-1}) we find
\begin{equation}
\mathcal{P}\sim 4\exp \left[ -2{\text{Im}}\mathcal{D}(\tau ^{+})\right] \sin
^{2}\left[ {\text{Re}}\mathcal{D}(\tau ^{+})\right] .  \label{P-DDP}
\end{equation}

We replace this expression by
\begin{equation}
\mathcal{P}\sim \frac{\sin ^{2}\left[ {\text{Re}}\mathcal{D}(\tau ^{+})%
\right] }{\cosh ^{2}\left[ {\text{Im}}\mathcal{D}(\tau ^{+})\right] }.
\label{P-DDP-sech}
\end{equation}%
There are several arguments in favour of this replacement. First of all, the
error we make when replacing Eq. (\ref{P-DDP}) with (\ref{P-DDP-sech}) is
comparable or smaller, and therefore negligible within the adiabatic limit,
where the DDP approximation itself is rigorously proved. Second, Eq. (\ref%
{P-DDP-sech}) is superior to Eq. (\ref{P-DDP}) because it does not violate
unitarity ($\mathcal{P}\leqq 1$), whereas Eq. (\ref{P-DDP}) does (albeit
only outside its range of validity). Lastly, such a replacement has already
been used \cite{Crothers} and shown to improve the accuracy.

\section{Conclusions\label{Sec-conclusion}}


We have examined the coherent excitation of two-state quantum system by
external pulsed field with Lorentzian temporal shape. The solution of the
problem can be expressed in terms of confluent Heun functions (CHF), but due
to some limitation of the analytical results for the CHF, analytical
approximation for the model is derived using DDP method.

\acknowledgments This work has been supported by the project QUANTNET -
European Reintegration Grant (ERG) - PERG07-GA-2010-268432.


\end{document}